\documentclass{article}
\usepackage{amsmath,graphicx,amssymb,bm,multirow,enumitem}
\usepackage[preprint]{spconf}

\setlist{nosep, leftmargin=14pt}

\title{Physics-Guided Dual-Domain Network with Attention-Based Fusion for Portable MRI Reconstruction}
\name{\textit{Efe Il\i cak}, Baris Imre, Chlo\'e Najac, Ruben van den Broek, Beatrice Lena, Andrew Webb, Marius Staring}
\address{Department of Radiology, Leiden University Medical Center, Leiden, Netherlands}

\toappear{\parbox{\textwidth}{
    \scriptsize 
    To appear in: {\it Proceedings of the 2026 IEEE International Symposium on Biomedical Imaging (ISBI), April 8-11, 2026, London, UK.} --- \copyright\ 2026 IEEE. Personal use of this material is permitted. Permission from IEEE must be obtained for all other uses, in any current or future media, including reprinting/republishing this material for advertising or promotional purposes, creating new collective works, for resale or redistribution to servers or lists, or reuse of any copyrighted component of this work in other works.
}}

\begin{document}
%
\maketitle

\begin{abstract}
Portable low-field magnetic resonance imaging (MRI) systems have gained renewed interest owing to their cost effectiveness and point-of-care imaging capabilities. Yet, portable MRI systems suffer from relatively low signal-to-noise ratio and limited hardware capabilities. While previous works have proposed the use of deep learning based reconstruction methods to improve low-field image quality, these operated only in the image-domain. Unlike other imaging modalities, MRI directly acquires data in the Fourier-domain (k-space), and exploiting both k-space and image-domain information can improve reconstruction quality. Here, we introduce DUN-DD, a novel physics-guided 3D network for portable MRI reconstruction, with parallel dual-domain branches whose outputs are combined together via an attention-based fusion network. To demonstrate the performance of the proposed method, we present \textit{in vivo} reconstructions obtained from both emulated datasets as well as images acquired with a 47mT Halbach-based portable MRI system. Our results show that DUN-DD outperforms state-of-the-art classical, data-driven, and physics-guided methods on both emulated and real portable MRI acquisitions.
\end{abstract}
\begin{keywords}
Low-field MRI, dual-domain learning, attention, deep unrolled network, cross-domain generalization
\end{keywords}
\section{Introduction}
\label{sec:intro}

Low-field MRI systems have recently gained renewed interest based on their cost effectiveness, operational safety, and portability \cite{Guallart-Naval2022}. These systems operate below 0.1T, and typically employ permanent magnets together wtih single-channel receiver coils, and electrical components powered by standard wall outlets \cite{Zhao2024}. Recent studies have demonstrated their diagnostic utility across diverse settings, from low-resource environments \cite{Jones2025} to point-of-care applications outside traditional clinical facilities \cite{Guallart-Naval2022}.

However achieving diagnostic-quality images within clinically feasible scan times with portable MRI systems remains a significant challenge due to limited signal-to-noise ratio, highly inhomogeneous main magnetic field, constrained gradient system performance, lack of multiple receiver arrays, and no RF shielded environment \cite{Liu2021,WebbTom2023}. To this end, recent studies have investigated deep-learning (DL) based methods in low-field MRI to improve the image quality while also improving scan efficiency. In \cite{Koonjoo2021}, the authors proposed a DL-based reconstruction to replace traditional image formation; whereas in \cite{Schlemper2019} and \cite{Zhou2022}, the authors proposed deep unrolled networks (DUNs) for reconstructing accelerated non-Cartesian acquisitions obtained with a 64mT scanner. More recently, a 3D DUN was developed for accelerated single-channel portable MRI reconstruction while utilizing additional prior information \cite{Ilicak2025}. Nonetheless, all of these methods operate solely in the image domain and do not leverage the complementary information present in k-space where MRI measurements are  acquired \cite{Cukur2025}.

To address this drawback, we introduce DUN-DD, a novel physics-guided dual-domain 3D network for portable MRI reconstruction. To this end, Fourier- and image-domain branches are processed in parallel, and the resulting representations of individual branches are fused together by an attention U-Net \cite{Oktay2018}. To demonstrate the performance of the proposed method, we present reconstructions obtained from both emulated \textit{in vivo} datasets, and real \textit{in vivo} acquisitions from a 47mT Halbach-based portable MRI system \cite{OReilly2021}.

\section{Methods}
\label{sec:methods}

\subsection{Problem Formulation}
Let $x\in \mathbb{C} $ be the ideal MR image, $y \in \mathbb{C}$ be the noisy k-space measurements, and $F_{\Omega}$ the forward operator with the sampling operator $\Omega$. The goal of MR image reconstruction is to recover $x$ from the measurements $y$, and this ill-posed problem can be cast as an optimization problem:
\begin{equation}
\underset{\bm{x}}{\text{arg min}} \|F_{\Omega}\bm{x}-\bm{y}\|_2^2  + \lambda R(\bm{x})
\label{Regularized Problem}
\end{equation}
Here, the first term represents data consistency, and the second term $R(\bm{x})$ denotes a regularization term, and $\lambda$ denotes the hyperparameter that balances data consistency and regularization. Traditionally, this problem can be solved via the gradient descent method: 
\begin{equation}
    \bm{x}^{i+1} =  \bm{x}^{i} - \alpha \nabla f(\bm{x}^{i})
\label{GradDesc}
\end{equation}
where $i$ denotes the iteration, $\alpha$ is the step size, $f$ is the objective function to be minimized. Alternatively, this problem can be solved via a physics-based model, by unrolling the steps of the iterative gradient descent algorithm to yield a DUN. Specifically, the iterations can be reformulated as \cite{Schlemper2019}:
\begin{equation}
    \begin{split}
        \bm{x}^{i+1} &= \bm{x}^i - \lambda^i  F_{\Omega}^H (F_{\Omega}\bm{x}^i - \bm{y} ) - N^{i}(\bm{x}^i)  \\
    \end{split}
\label{Proposed}
\end{equation}
where $N^{i}$ denotes the learnable network module, and both $\lambda^i$ and $N^{i}$ are the trainable parameters for each iteration block. 

\subsection{Datasets}
Supervised methods yield high reconstruction accuracy, but they require large fully-sampled raw datasets \cite{Cukur2025}. However, no such large datasets are publicly available for low-field MRI systems. To address this, we propose a framework \cite{Ilicak2025} for emulating low-field k-space measurements from high-field acquisitions. Using the fastMRI brain dataset \cite{Zbontar2018}, we emulated datasets containing 400 training, 100 validation and 80 test subjects from fluid-attenuated inversion recovery (FLAIR), T1-weighted, contrast-enhanced T1-weighted, and T2-weighted acquisitions.

Additionally, to evaluate the performance of the proposed DUN-DD model, inversion recovery T1- (IR-T1-w), proton density- (PD-w), and T2-weighted brain images were acquired from five healthy volunteers using a prototype 47mT portable MRI system \cite{OReilly2021}. Imaging was performed with 3D turbo spin-echo protocols at a resolution of $1.5\times1.5\times5$ mm$^3$ with a matrix size of $150\times136\times38$. All scans were approved by the local ethics committee, and written informed consent was obtained from all participants. These acquisitions were not included in the network trainings. Moreover, due to the relaxation time and protocol differences between the field strengths, these acquisitions exhibit different image contrasts compared to training data, thereby representing a domain shift. Consequently, the reconstruction performance on 47mT acquisitions also reflect the cross-domain generalization capabilities of the networks.  

\subsection{Implementation Details}
\begin{figure}
    \centering
    \includegraphics[width=1\linewidth]{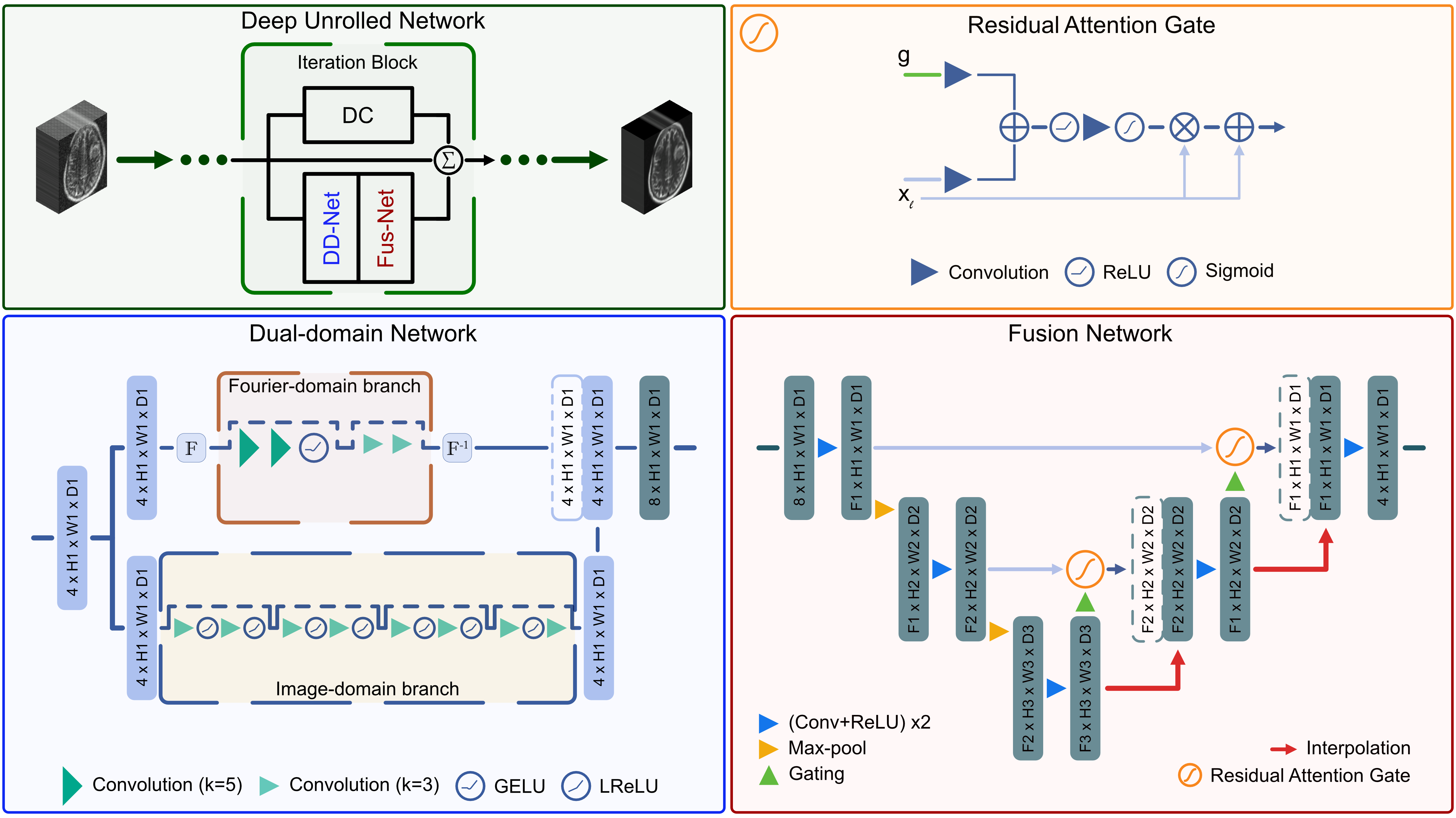}
    \caption{Architecture of the proposed DUN-DD network. Each iteration includes parallel Fourier- and image-domain branches whose outputs are fused via an attention-based network, together with data consistency (DC) projections.}
    \label{fig:network}
\end{figure}

We implemented the unrolled network using 5 iteration blocks, with a total of 3.53M trainable parameters. Each iteration block is comprised of the previous estimate, a soft data consistency (DC) projection, and a dual-domain convolutional neural network (CNN). In each CNN, a parallel stream of Fourier- and image-domain branches reconstruct the images separately, and the resulting representations are concatenated and fused via a novel residual attention U-Net. To exploit the inherent complex-conjugate symmetry of k-space, we augmented the input tensor with its conjugate-symmetric counterpart using the virtual conjugate coil method, and split the resulting tensor into real and imaginary parts, yielding a 4-channel input for both branches \cite{Ilicak2025}. The network architecture and its components are depicted in Figure \ref{fig:network}.


\section{Experiments and Results}
\label{sec:exp}
To evaluate DUN-DD, we retrospectively undersampled the emulated datasets as well as 47mT acquisitions at acceleration rate $R=2$. Undersampling was achieved via two-dimensional variable density masks with 12\% fully-sampled k-space center. All models were trained using a combination of Huber and a perceptual loss \cite{Ghodrati2019}, and a new random sampling mask was generated independently for each subject at every training epoch. Differences from DUN-DD were tested for statistical significance using the Wilcoxon signed-rank test, with $^*$ denoting $p<0.01$.

\subsection{Ablation Study}
To assess the contribution of components in DUN-DD, we conducted ablation experiments on the emulated datasets. We evaluated three ablated variants: \textbf{DUN-kSpc}, Fourier-domain only; \textbf{DUN-Img}, image-domain only; and \textbf{DUN-Avg}, dual-domain with naive averaging instead of attention-based fusion. As listed in Table \ref{tab:Ablation}, the proposed attention-incorporated dual-domain DUN-DD achieved the best performance.
\begin{table}[h]
    \centering
    \small
    \begin{tabular}{|c|c|c|}
        \hline
        Method & PSNR & SSIM  \\
        \hline
        DUN-kSpc     & 27.24±2.78$^*$       & 0.83±0.05$^*$ \\
        DUN-Img     & 30.13±2.68$^*$        &  0.89±0.05$^*$   \\
        DUN-Avg     & 29.88±2.66$^*$        & 0.89±0.05$^*$ \\
        DUN-DD      &\textbf{30.20±2.69}{ } & \textbf{0.89±0.08}{ }    \\
        \hline
    \end{tabular}
    \caption{Ablation results on the emulated low-field datasets at $R=2$. Bold indicates the best performance.}
    \label{tab:Ablation}
\end{table}

\subsection{Emulated Dataset Results}
The proposed DUN-DD model was compared to state-of-the-art classical and DL-based reconstructions to showcase its performance. These include: \textbf{BM4D}, classical volumetric denoising method based on block matching with 4D filtering \cite{Maggioni2013}; \textbf{CS-TV}, total variation compressed sensing reconstruction with automatic regularization weight selection \cite{Ilicak2023}; \textbf{U-Net} architecture with 5-levels and 3D kernels; \textbf{ViT} architecture with patch size of $4\times9\times9$, MLP dimension of 768, hidden layer dimension of 384, 12 layers, 8 heads; \textbf{ISTA-Net}, a 2D unrolled network based on iterative shrinkage-thresholding algorithm \cite{Zhang2018}. 
\begin{table}[tb!]
    \centering
    \small
    \begin{tabular}{|c|c|c|c|}
        \hline
        Method & PSNR & SSIM & Infer. (s) \\
        \hline
        BM4D        & 26.16±2.16$^*$                & 0.79±0.05$^*$             & 5.32±0.25\\
        CS-TV       & 26.35±1.96$^*$                & 0.79±0.05$^*$             & 5.75±0.51\\
        U-Net       & 27.72±1.94$^*$                & 0.85±0.05$^*$             & 0.04±0.01\\
        ViT         & 25.92±1.77$^*$                & 0.81±0.05$^*$             & 0.01±0.01\\
        ISTA-Net    & \underline{28.66±2.38}$^*$    & \underline{0.86±0.05}$^*$ & 0.98±0.03\\
        DUN-DD      &\textbf{30.20±2.69}{ }         & \textbf{0.89±0.05}{ }     & 0.65±0.06\\
        \hline
    \end{tabular}
    \caption{Reconstruction performance of competing methods on the emulated low-field datasets at $R=2$.  Bold and underlined indicate best and second-best performance.}
    \label{tab:Testset}
\end{table}

The quantitative performance and inference times of the reconstruction methods are listed in Table \ref{tab:Testset}. Overall, we observe that physics-guided models (ISTA-Net, DUN-DD) outperform both the classical and data-driven models, with the proposed DUN-DD achieving the highest PSNR and SSIM scores.

\subsection{47mT Portable MRI Results}
\begin{figure}
    \centering
    \includegraphics[width=1\linewidth]{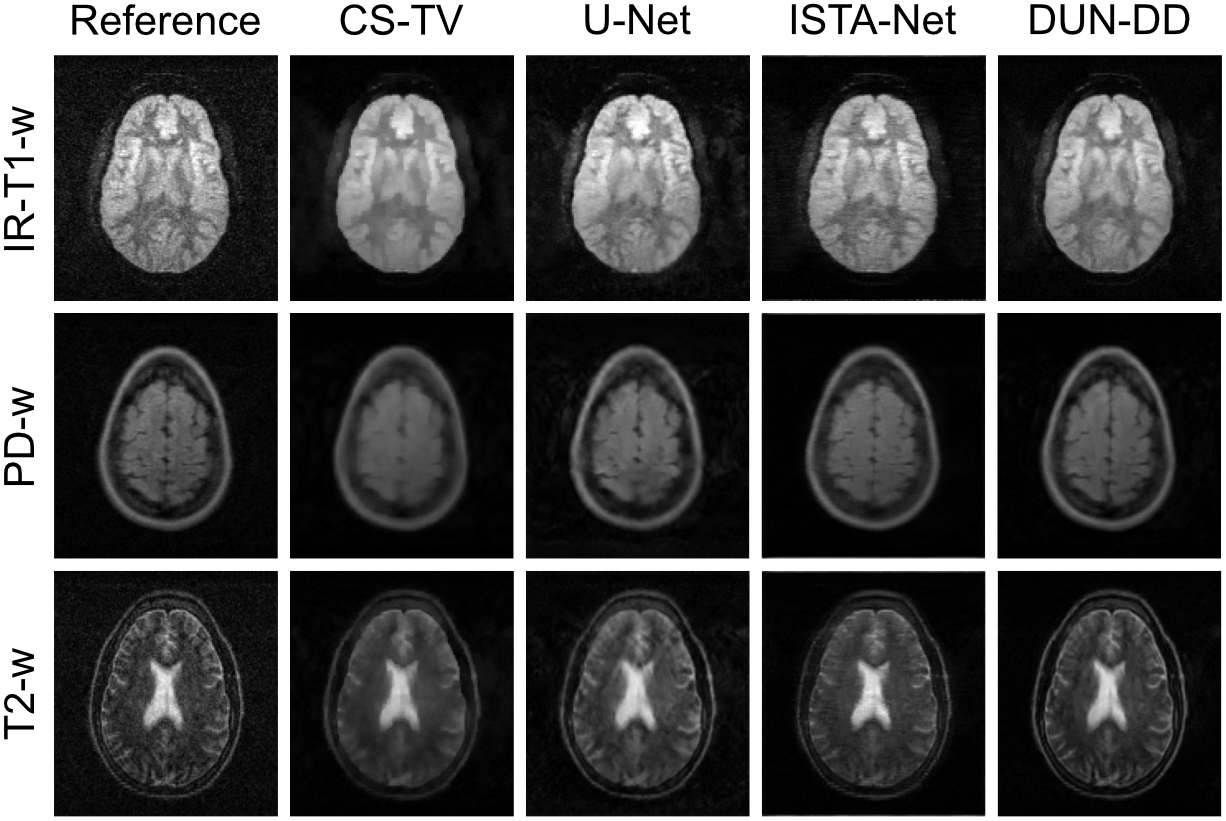}
    \caption{Representative reconstructions at $R=2$ for the 47mT portable MRI acquisitions, together with fully-sampled references. CS-TV exhibits excessive blurring, and U-Net shows residual aliasing artifacts. ISTA-Net reduces artifacts but the proposed DUN-DD achieves the better performance with superior artifact suppression and sharper details.}
    \label{fig:Halbach}
\end{figure}

The proposed DUN-DD, and the best performing classical (CS-TV), data-driven (U-Net), and physics-guided (ISTA-Net) methods were then tested on 47mT MRI acquisitions at $R=2$. Figure \ref{fig:Halbach} illustrates representative reconstructions, together with fully-sampled references, and Table \ref{tab:Halbach} lists the quantitative results. 
\begin{table}[b]
    \centering
    \small
    \begin{tabular}{|c|c|c|c|}
        \hline
         Contrast & Method & PSNR & SSIM\\
         \hline
                    & CS-TV     & 25.11±1.12                & 0.83±0.01  \\
         IR-T1-w    & U-Net     & 24.09±0.98                & 0.85±0.02  \\
                    & ISTA-Net  & \underline{25.74±1.16}    & \underline{0.87±0.02}  \\
                     & DUN-DD   & \textbf{25.98±1.46}       & \textbf{0.88±0.02}    \\
        \hline
                    & CS-TV     & 29.43±2.26                & 0.87±0.01  \\
         PD-w       & U-Net     & 28.24±2.26                &  0.89±0.01  \\
                    & ISTA-Net  & \underline{30.26±2.08}    & \underline{0.90±0.01}   \\
                    & DUN-DD    & \textbf{31.91±2.76}       & \textbf{0.93±0.02}   \\
        \hline
                    & CS-TV     & 25.79±1.47                & 0.75±0.01  \\
         T2-w       & U-Net     & 25.23±1.29                & 0.82±0.02  \\
                    & ISTA-Net  & \underline{26.44±1.33}  & \underline{0.83±0.01}   \\
                    & DUN-DD    & \textbf{27.29±1.68}       & \textbf{0.87±0.02}   \\
         \hline
    \end{tabular}
    \caption{Reconstruction performance at $R=2$ across image contrasts from the 47mT portable MRI system. Bold and underlined indicate best and second-best performance.}
    \label{tab:Halbach}
\end{table}
Overall, we observe that DUN-DD provides robust reconstructions with improved artifact suppression and sharper details, and it consistently ranks as the top performer in terms of PSNR and SSIM metrics.

\section{Discussion and Conclusion}
We presented DUN-DD, a physics-guided network for portable MR image reconstruction. DUN-DD improves upon the single-domain models by reconstructing the data via parallel Fourier- and image-domain branches, and it combines these representations by a novel fusion network based on residual attention U-Net architecture. The network was trained only on emulated low-field images, and was tested on both emulated and real 47mT portable MRI acquisitions. 

The results across both datasets indicate that DUN-DD is capable of improving the reconstruction quality compared to state-of-the-art methods. Furthermore, it achieves robust performance on real portable MRI acquisitions with unseen image contrasts, highlighting its strong cross-domain generalization capabilities.



\section{Acknowledgments}
\label{sec:acknowledgments}
This work was supported by the European Innovation Council (EIC-Transition 101136407).

\bibliographystyle{IEEEbib}
\bibliography{refs}

\end{document}